%% file: base.tex
\documentclass[sigconf, authorversion, nonacm]{acmart}

\usepackage{pifont}%

\usepackage{xcolor}
\usepackage{colortbl}

\usepackage{multirow}
\usepackage{arydshln}
\usepackage{threeparttable}  
\usepackage{tcolorbox}

\AtBeginDocument{%
  }

\begin{document}

\title{An LLM-Tool Compiler for Fused Parallel Function Calling}

\author{Simranjit Singh}
\affiliation{%
  \institution{Microsoft Corporation}
  \city{Silicon Valley Campus}
  \state{CA}
  \country{USA}
}
\email{simsingh@microsoft.com}

\author{Andreas Karatzas}
\affiliation{%
  \institution{Southern Illinois University}
  \city{Carbondale}
  \state{IL}
  \country{USA}
}
\email{andreas.karatzas@siu.edu}

\author{Michael Fore}
\affiliation{%
  \institution{Microsoft Corporation}
  \city{Reston}
  \state{VA}
  \country{USA}
}
\email{mifore@microsoft.com}

\author{Iraklis Anagnostopoulos}
\affiliation{%
  \institution{Southern Illinois University}
  \city{Carbondale}
  \state{IL}
  \country{USA}
}
\email{iraklis.anagno@siu.edu}

\author{Dimitrios Stamoulis}
\affiliation{%
  \institution{Microsoft Corporation}
  \city{Redmond}
  \state{WA}
  \country{USA}
}
\email{stamoulis.dimitrios@microsoft.com}


\begin{abstract}
State-of-the-art sequential reasoning in Large Language Models (LLMs) has expanded the capabilities of Copilots beyond conversational tasks to complex function calling, managing thousands of API calls. However, the tendency of compositional prompting to segment tasks into multiple steps, each requiring a round-trip to the GPT APIs, leads to increased system latency and costs. Although recent advancements in parallel function calling have improved tool execution per API call, they may necessitate more detailed in-context instructions and task breakdown at the prompt level, resulting in higher engineering and production costs. Inspired by the hardware design principles of multiply-add (MAD) operations, which fuse multiple arithmetic operations into a single task from the compiler’s perspective, we propose \texttt{LLM-Tool Compiler}, which selectively \textit{fuses similar} types of tool operations under a single function at runtime, presenting them as a unified task to the LLM. This selective fusion inherently enhances parallelization and efficiency. Benchmarked on a large-scale Copilot platform, \texttt{LLM-Tool Compiler} achieves up to four times more parallel calls than existing methods, reducing token costs and latency by up to 40\% and 12\%, respectively.
\end{abstract}

\begin{teaserfigure}
  \includegraphics[width=\textwidth]{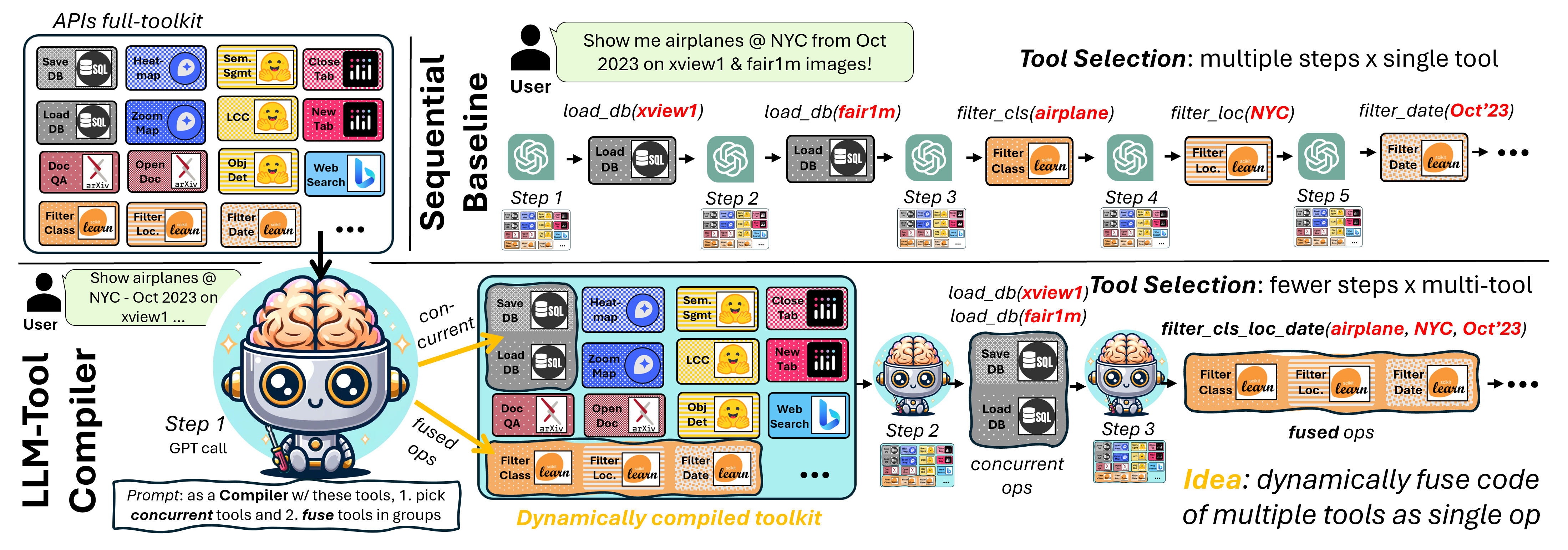}
  \caption{Fused parallel function calling with \texttt{LLM-Tool Compiler}. For each user query, the compiler dynamically identifies tools to execute concurrently and/or \textbf{\textit{fuses}} similar functions to single operations. By presenting them as unified tasks to GPT, our approach inherently enhances parallelization and overall efficiency. Results show that \texttt{LLM-Tool Compiler} increases parallel calls by up to five times when integrated with various prompting schemes.}
  \Description{An overview of our proposed Tool-LLM Compiler method.}
  \label{fig:teaser}
\end{teaserfigure}


\maketitle

\section{Introduction}

Recent advances in Large Language Models (LLMs) have significantly enhanced their reasoning capabilities, positioning them as novel Copilots for diverse applications in robotics, AR/VR, and geospatial tasks~\cite{OpenEQA2023}. These models proficiently manage APIs, UI-Web interfaces, operating systems, apps, and SQL backends~\cite{koh2024visualwebarena}. However, this expanded functionality increasingly strains the entire system stack, from cloud endpoints to local execution devices, introducing substantial hardware overhead~\cite{fore2024geckopt}, yet offering numerous opportunities to improve metrics such as latency, memory usage, and energy efficiency~\cite{yuan2024roof}. The growing necessity for system-level optimizations signifies a paradigm shift toward system-oriented designs in the development of Copilots, a shift that is increasingly recognized as essential by our systems engineering community~\cite{kim2024compiler}.

A novel advancement enabling impressive Copilot performance is robust reasoning through compositional and sequential prompting, with techniques such as Chain-of-Thought~\cite{wei2023chainofthought}, ReAct~\cite{yao2023react}, and Tree-of-Thought~\cite{yao2023tree} enhancing function-calling benchmarks substantially. However, their inclination to decompose tasks into single steps~\cite{lu2023chameleon}, each necessitating a separate GPT OpenAI API call, introduces significant bottlenecks, thereby increasing token costs and system latency (Figure 1, top). Recent developments in parallel function calling aim at selecting and executing multiple tools within a single API response (Figure 1, right)~\cite{oaiparallel}. Notably, OpenAI's introduction of that very feature in their recent GPT Turbo release exemplifies this advancement\footnote{We use the term \textit{parallel} in the same sense as the OpenAI API documentation, which refers to returning multiple tools per GPT call to ``\textit{reduce round trips with the API}''~\cite{oaiparallel}. This should not be confused with the actual execution of the tools, which may not run as parallel processes. To avoid ambiguity, we define \textit{concurrent} as the simultaneous execution of multiple tools, while \textit{parallel} specifically denotes the capability of a single GPT API call to return multiple tools.}. Despite improved performance, these methods still frequently require nuanced prompting to effectively guide the agent on how to group functions. For instance, few-shot techniques that provide in-context previous examples of multi-tool execution that encourage effective function selection often necessitate complex, RAG-based \cite{lewis2021retrievalaugmented} or intent-based \cite{fore2024geckopt} dynamic prompting schemes.

A key insight in our work is that in realistic LLM workloads, data often exhibits significant temporal similarity, with sequences of similar tools typically executed in succession. Consider as an example an LLM-enabled geospatial analysis platform in which a user may succinctly ask, ``Show me satellite images around Newark, NJ from October.'' This query naturally involves filtering data first by location and then by date. If the LLM could interpret these sequential tasks as a single function -- though the system still processes them with their respective code internally -- the agent would, in essence, select one integrated tool. This approach inherently reduces the frequency of multiple single-tool API calls, streamlining the process and enhancing efficiency.

Drawing inspiration from hardware systems, akin to the principles of the multiply-add (MAD) operation that fuses different arithmetic steps into a single task from the compiler’s perspective, we introduce a GPT-driven tool-compilation scheme, namely \texttt{LLM-Tool Compiler}. Prior to any tool execution, our compiler identifies groups of needed tools and compiles them into fused operations. Without modifying the function-calling API, tool selection proceeds as usual, but with a significant enhancement: at the LLM level, the agent is presented with and selects from an updated and more granular list of functions, thereby inherently achieving a higher parallelization rate in function calling. Upon selection, the compiler then ``maps'' the fused operations back to their corresponding code per tool. Through extensive evaluation on a state-of-the-art geospatial Copilot framework~\cite{singh2024geoengine}, we demonstrate that our approach enhances LLM performance, reducing token costs and latency by 12\% and improving parallelization by up to four times across various LLM models and prompting techniques.

\section{Related Work}

The nascent landscape of LLM optimization strategies encompasses a diverse array of approaches, ranging from model-level improvements~\cite{kim2024squeezellm, frantar2023gptq, lin2024awq, dettmers2023spqr, kwon2022fast, meta-llama3} to system-level enhancements~\cite{NVIDIA-TensorRT-LLM, HuggingFace-TextGenInference, kwon2023efficient}. As token length directly correlates with system overhead, significant advancements in LLM efficiency have been achieved through input compression by reducing the cost associated with longer prompts. Methods such as DYNAICL~\cite{zhou2023efficient}, Selective Context~\cite{li2023compressing}, and LLMLingua~\cite{jiang2023llmlingua} dynamically remove non-essential elements from prompts, streamlining input without compromising information quality. RECOMP~\cite{xu2023recomp} and SemanticCompression~\cite{fei2023extending} condense prompts into concise summaries that maintain essential semantic content. All these approaches are orthogonal to our approach and could be flexibly integrated into the \texttt{LLM-Tool Compiler}, as we focus on the tool-calling agent level.

Recent strategies have been developed to enhance the decoding efficiency of LLMs, such as InContext Learning (ICL)\cite{dong2023survey} and Chain-of-Thought (CoT)\cite{wei2023chainofthought}, which incorporate multiple examples and reasoning steps within prompts. Moreover, methods like RAG~\cite{lewis2021retrievalaugmented}, FLARE~\cite{chevalier2023adapting}, and intent-based function-calling~\cite{fore2024geckopt} streamline tool selection for LLMs, leading to significant cost reductions. While these advancements are crucial in optimizing agent efficiency, they have primarily focused on agent performance metrics and only indirectly on hardware-related optimizations. Our work shares this motivation but also explicitly profiles and reports hardware metrics, including system latency and token cost, on top of a large-scale cloud Copilot framework with hundreds of GPT endpoints, terabytes of data, and long-horizon complex tasks. This provides a more comprehensive approach that reflects industry-scale hardware and real-world applications.

Highlighting the growing demand for more efficient agents, OpenAI introduced parallel function calling support as part of their latest GPT release (\texttt{Turbo}) setting state-of-the-art performance~\cite{berkeley-function-calling-leaderboard}. Our results show that our \texttt{LLM-Tool Compiler} further improves upon this strong baseline. Skeleton-of-Thought (SoT)\cite{ning2024skeletonofthought} and its enhancements like SGD\cite{jin2024adaptive} and APAR~\cite{liu2024apar} restructure the output content into a manageable format conducive to batch processing and parallel decoding, significantly reducing generation latency while enriching the diversity and quality of responses. SGLang~\cite{zheng2023efficiently} introduces a domain-specific programming language tailored for LLM operations, promoting batch inference and efficient resource sharing. However, these approaches often assume embarrassingly parallel or inherently interdependent tasks, as highlighted by the authors in~\cite{kim2024compiler}. Concurrently with our submission, \texttt{LLMCompiler}~\cite{kim2024compiler} introduced a compiler-based methodology for parallel function-level tool selection. However, their approach is meant more as an end-to-end prompting solution to replace existing schemes like ReAct, hence necessitating extensive user-specified prompting. In contrast, our framework operates directly at the toolset level and is nearly agnostic to the prompting scheme, allowing us to reuse the existing reasoning logic, thus minimizing engineering overhead.

\section{Methodology}

To illustrate the intuition behind the \texttt{LLM-Tool Compiler}, consider the task depicted in Figure~\ref{fig:teaser}, where a user asks, ``\textit{Show me airplanes at NYC from October 2023 on xview1 and FAIR1M images}.'' In the fully sequential baseline, this query is completed with the agent calling and executing one tool per step, i.e., per GPT API call (Figure~\ref{fig:teaser}, top), where it is evident that several tools could be executed simultaneously. Indeed, in this particular example, OpenAI's latest release achieves parallelization by executing the load operations simultaneously. However, there is considerable potential for further improvement, especially for the subsequent data operations: in fact, as we discuss later in our Results, only a quarter of data filtering operations are currently executed in parallel by the GPT baseline.

We hypothesize that the limited parallelization of data operations may be partly due to the high diversity among 30 different data tools. Although each function definition is available to the agent, nuanced interdependencies among these tools might not be apparent. One potential approach could involve employing overly detailed prompting instructions and meticulously crafted tool-specific directions. However, we consider this approach impractical for real-world, large-scale Copilot systems~\cite{koh2024visualwebarena}, where updating production code to reflect new tool aspects and interdependencies would be cumbersome. In fact, as we observe in practice, even more detailed prompting instructions to consider more data operations in a single step do not ensure that the agent will execute multi-tool operations, as reported by experiences from other practitioners~\cite{oaiparallelforum}.

The critical question is how to dynamically determine, for each new user prompt, which tasks to parallelize and orchestrate their execution without significantly altering the preexisting function-calling process. To achieve this, we propose decoupling the "compilation" step as much as possible in an agent-agnostic manner, by introducing a dedicated GPT-driven module. Before any tool execution and given the current query and the full API tool-set, this module identifies groupings of similar tools that can be invoked together. This design choice is deliberate, as it confines the prompting effort and tool compilation within that module as a separate function call, without replacing or overly modifying existing agent baselines. As depicted in Figure~\ref{fig:teaser} (bottom), the \texttt{LLM-Tool Compiler} accomplishes this through two primary components: (i) a fuser that identifies relevant tools and aggregates them into a single function, dynamically updating the tool-set visible to the LLM agent; and (ii) an executor that, during the subsequent function-calling process, monitors the tool selection by the agent. If a fused operation is selected, the executor is responsible for decomposing and executing their respective code blocks and dependencies:

\textbf{Fuser}. This lightweight, GPT-driven module runs prior to the main agent API calls and is responsible for identifying and assembling groups of tools that can be fused into more granular tasks. To enable this, we leverage user-supplied predefined prompts that instruct the module on which tools are associated with different types of tasks via intent-based in-context prompting~\cite{fore2024geckopt}. Specifically, we adapt intent-based instructions to consider the aspects for parallel function calling, where we also guide how various groups of tasks can be parallelized. Referring to our earlier example query of load-filter-plot from Figure~\ref{fig:teaser}, this involves clarifying to the fuser that certain filtering operations (\textit{e.g.}, by date or coordinates) are candidates for grouping. We provide an example system prompt, abbreviated for brevity.

The fuser module invokes the GPT API with the user query, the system prompt, and the complete list of API tool descriptions. We force the GPT API to return a predefined JSON schema that contains the tool groupings needed to answer the user query. This involves calling the GPT API using the \texttt{tool\_choice: "none"} API flag. Based on the GPT output message (in JSON format), we perform a sanity check to ensure that (i) the selected groupings correspond to existing function names, and (ii) they are associated with the correct tool types via a simple lookup table. Subsequently, for each fused list, we programmatically aggregate the tool definitions into unified tasks, which includes all variable names, descriptions, etc. This step simply consolidates all variables (\textit{i.e.}, \texttt{name}, \texttt{description}, \texttt{parameters} entries in function-call API) for each fused tool into unified arguments lists. Each fused tool is then incorporated into the tool-set in place of the sublist of tools it replaces. From the agent’s perspective, this process is seamless, as it only interacts with an updated tool set.

\begin{tcolorbox}[title=\textit{Fuser} prompt via intent-based in-context examples, colback=gray!20, colframe=gray!75, rounded corners, sharp corners=northeast, sharp corners=southwest]
\footnotesize
\texttt{As a Compiler with access to tools for geospatial tasks [..] \\
~\\
\textbf{Intents}: 1. load->filter->plot; 2. UI/web; 3. Doc retrieval\\
~\\
Given the examples below and *without* solving for the query, reason about which tools will be *likely* needed to complete the task, then compile lists of tools for each subset of similar ops (e.g., filter, load, map, doc ops) without calling any arguments.\\
~\\
\textbf{User Query}: \{question\}\\
\textbf{In-context examples}: \\
~\\
\textbf{Example 1}: Plot on the map the fmow images in Europe from June 2012
Thought 1: The user is asking for images of a specific dataset, so I need to load [..]\\
Thought 2: [..] images for a month (therefore date range) and a continent (therefore location)\\
"Action": To complete the task I would call [..]\\
- load\_db(..load the fmow images..)\\
- filter\_loc(..filter in Europe...)\\
- filter\_date(..filter from June 2012..)\\
Answer: .. \\
~\\
\textbf{Example 2}: [..] \\
------\\
Answer the format of json: \{'load\_ops': [..], 'filter\_ops': [..], ..\}. Think step by step 
}
\end{tcolorbox}

\begin{table*}
  \caption{System and agent performance for GPT-3.5 Turbo and GPT-4 Turbo and various prompting techniques.}
  \label{tab:latency_results}
  \begin{tabular}{lccccccc}
        \toprule
        \multirow{2}{*}{Model} & \multirow{2}{*}{Compiler} & Success & Avg Tokens & Token & Avg Time & \multirow{2}{*}{Speedup $\uparrow$} \\
         & & Rate (\%) $\uparrow$ & /Task $\downarrow$ & Reduction $\uparrow$ & /Task (s) $\downarrow$ &  \\
        \midrule
        \textbf{\textit{GPT-3.5 Turbo (0125)}} & & & & & & & \\
        \multirow{3}{*}{CoT - Zero-Shot} & $^\dagger$Baseline: In-context & 65.73$\pm$0.45 & 25.40$\pm$0.02 & --  & 8.11$\pm$0.01 & -- \\
         & $^\dagger$Baseline: In-context + OAI Parallel & 65.93$\pm$0.49 & 25.15$\pm$0.02 & 1.01 $\times$  & 7.44$\pm$0.01 & 1.09 $\times$ \\
         & \cellcolor{teal!15}\texttt{LLM-Tool Compiler} & \cellcolor{teal!15}58.59$\pm$0.57 & \cellcolor{teal!15}21.17$\pm$0.01 & \cellcolor{teal!15}1.20 $\times$  & \cellcolor{teal!15}5.84$\pm$0.00 & \cellcolor{teal!15}1.39 $\times$ \\
        \hdashline
        \multirow{3}{*}{CoT - Few-Shot} & $^\dagger$Baseline: In-context & 67.87$\pm$0.20 & 30.36$\pm$0.01 & --  & 8.09$\pm$0.00 & -- \\
         & $^\dagger$Baseline: In-context + OAI Parallel & 67.17$\pm$0.57 & 30.22$\pm$0.02 & 1.00 $\times$  & 7.49$\pm$0.01 & 1.08 $\times$ \\
         & \cellcolor{teal!15}\texttt{LLM-Tool Compiler} & \cellcolor{teal!15}59.55$\pm$0.20 & \cellcolor{teal!15}25.19$\pm$0.01 & \cellcolor{teal!15}1.21 $\times$  & \cellcolor{teal!15}5.97$\pm$0.01 & \cellcolor{teal!15}1.36 $\times$ \\
        \hdashline
        \multirow{3}{*}{ReAct - Zero-Shot} & $^\dagger$Baseline: In-context & 64.81$\pm$0.63 & 27.52$\pm$0.14 & --  & 8.13$\pm$0.02 & -- \\
         & $^\dagger$Baseline: In-context + OAI Parallel & 64.98$\pm$0.20 & 23.92$\pm$0.09 & 1.15 $\times$  & 6.97$\pm$0.02 & 1.17 $\times$ \\
         & \cellcolor{teal!15}\texttt{LLM-Tool Compiler} & \cellcolor{teal!15}59.33$\pm$0.67 & \cellcolor{teal!15}21.06$\pm$0.02 & \cellcolor{teal!15}1.31 $\times$  & \cellcolor{teal!15}5.12$\pm$0.02 & \cellcolor{teal!15}1.59 $\times$ \\
        \hdashline
        \multirow{3}{*}{ReAct - Few-Shot} & $^\dagger$Baseline: In-context & 73.59$\pm$0.13 & 39.44$\pm$0.01 & --  & 8.19$\pm$0.01 & -- \\
         & $^\dagger$Baseline: In-context + OAI Parallel & 73.20$\pm$0.19 & 35.24$\pm$0.05 & 1.12 $\times$  & 7.15$\pm$0.01 & 1.15 $\times$ \\
         & \cellcolor{teal!15}\texttt{LLM-Tool Compiler} & \cellcolor{teal!15}66.12$\pm$1.05 & \cellcolor{teal!15}29.29$\pm$0.12 & \cellcolor{teal!15}1.35 $\times$  & \cellcolor{teal!15}5.30$\pm$0.04 & \cellcolor{teal!15}1.54 $\times$ \\
        \midrule
        \textbf{\textit{GPT-4 Turbo (0125)}} & & & & & & & \\
        \multirow{3}{*}{CoT - Zero-Shot} & $^\dagger$Baseline: In-context & 84.39$\pm$0.14 & 29.52$\pm$0.01 & --  & 8.98$\pm$0.00 & -- \\
         & $^\dagger$Baseline: In-context + OAI Parallel & 84.73$\pm$0.21 & 24.49$\pm$0.03 & 1.21 $\times$  & 8.44$\pm$0.00 & 1.06 $\times$ \\
         & \cellcolor{teal!15}\texttt{LLM-Tool Compiler} & \cellcolor{teal!15}84.15$\pm$0.20 & \cellcolor{teal!15}21.30$\pm$0.01 & \cellcolor{teal!15}1.39 $\times$  & \cellcolor{teal!15}7.40$\pm$0.00 & \cellcolor{teal!15}1.21 $\times$ \\
        \hdashline
        \multirow{3}{*}{CoT - Few-Shot} & $^\dagger$Baseline: In-context & 83.15$\pm$0.22 & 33.75$\pm$0.04 & --  & 8.83$\pm$0.02 & -- \\
         & $^\dagger$Baseline: In-context + OAI Parallel & 83.05$\pm$0.46 & 26.50$\pm$0.00 & 1.27 $\times$  & 8.19$\pm$0.03 & 1.08 $\times$ \\
         & \cellcolor{teal!15}\texttt{LLM-Tool Compiler} & \cellcolor{teal!15}82.51$\pm$0.03 & \cellcolor{teal!15}23.89$\pm$0.00 & \cellcolor{teal!15}1.41 $\times$  & \cellcolor{teal!15}7.40$\pm$0.00 & \cellcolor{teal!15}1.19 $\times$ \\
        \hdashline
        \multirow{3}{*}{ReAct - Zero-Shot} & $^\dagger$Baseline: In-context & 82.02$\pm$0.27 & 32.55$\pm$0.00 & --  & 8.97$\pm$0.00 & -- \\
         & $^\dagger$Baseline: In-context + OAI Parallel & 82.31$\pm$1.04 & 27.50$\pm$0.00 & 1.18 $\times$  & 8.42$\pm$0.00 & 1.06 $\times$ \\
         & \cellcolor{teal!15}\texttt{LLM-Tool Compiler} & \cellcolor{teal!15}81.68$\pm$0.31 & \cellcolor{teal!15}24.35$\pm$0.01 & \cellcolor{teal!15}1.34 $\times$  & \cellcolor{teal!15}7.38$\pm$0.00 & \cellcolor{teal!15}1.21 $\times$ \\
        \hdashline
        \multirow{3}{*}{ReAct - Few-Shot} & $^\dagger$Baseline: In-context & 84.26$\pm$0.29 & 39.18$\pm$0.01 & --  & 8.94$\pm$0.00 & -- \\
         & $^\dagger$Baseline: In-context + OAI Parallel & 84.29$\pm$0.42 & 33.72$\pm$0.01 & 1.16 $\times$  & 8.41$\pm$0.01 & 1.06 $\times$ \\
         & \cellcolor{teal!15}\texttt{LLM-Tool Compiler} & \cellcolor{teal!15}82.68$\pm$0.27 & \cellcolor{teal!15}29.36$\pm$0.01 & \cellcolor{teal!15}1.33 $\times$  & \cellcolor{teal!15}7.38$\pm$0.00 & \cellcolor{teal!15}1.21 $\times$ \\
        \bottomrule
\end{tabular}
\end{table*}

\textbf{Function calling.} It is critical to note that, aside from the fuser initial step, there are no further changes to the function-calling process and the reasoning scheme. The system prompts, underlying tool definitions, and prompting strategies remain unchanged, as the only modification involves aggregating groups of tools into more granular multi-tool function choices. In contrast to compiler methods that directly replace tool-selection schemes like ReAct~\cite{kim2024compiler}, our method can be added on top of any technique. This compatibility is demonstrated in our results, which apply to both zero and few-shot paradigms for Chain-of-Thought and React. To illustrate, Step 3 in Figure 1 shows that following the same query example as before, the GPT agent sees an updated tool-set, but the function definitions are the original ones, allowing it to select the consolidated filter-by-date-location tool to complete the task.

\textbf{Executor.} The executor module actively monitors the GPT tool selection output. When a fused function name is selected, it is responsible for de-fusing the operation into its constituent sub-tasks. Referring back to the example in Figure 1, if the agent selects the granular tool 'filter-by-date-by-location,' the executor seamlessly manages the execution of both 'filter-by-date' and 'filter-by-location.' This operation is programmatically determined based on the tool definitions. For independent tasks, such as load operations, the executor executes them concurrently; for tasks that modify the same dataset -- like sequentially filtering by dates and then by location -- it processes them sequentially. As demonstrated in our ablation studies, the primary speed-up results from the reduction in the number of steps through multi-tool groupings, rather than from faster execution of individual steps.

\textbf{Error handling.} Given that the fuser is a fully GPT-driven module, our method handles error cases at two levels. Initially, the fuser is instructed to return the JSON of fused tasks only when confident; otherwise, it provides an empty list, in which case the toolset remains unchanged and the agent proceeds as usual. Moreover, the agent is prompted to restart if it encounters issues during the process after a few failed attempts. Upon reset, we bypass the fuser API call and repeat the operation with the full toolset. This dynamic adaptability is crucial for rectifying inaccuracies in tool selection and enables the system to address issues in real-time.

\input{tables.tex}

\begin{table*}
  \caption{Benchmark parallelization analysis: Evaluating the parallelization rate (\%) achieved per method with respect to the overall number of parallelizable function calls.}
  \label{tab:parallelization_analysis}    
  \begin{tabular}{lccccc}
        \toprule
        \multirow{2}{*}{Model} & \multirow{2}{*}{Compiler} & \multicolumn{2}{c}{Parallelization - Load Ops (\%)} & \multicolumn{2}{c}{Parallelization - Filter Ops (\%)}  \\
        & & \textbf{\textit{GPT-3.5}} & \textbf{\textit{GPT-4}} & \textbf{\textit{GPT-3.5}} & \textbf{\textit{GPT-4}}\\
        \midrule
        \multirow{2}{*}{CoT - Zero-Shot} & $^\dagger$Baseline: In-context + OAI Parallel &  1.02$\pm$0.08 & 99.85$\pm$0.04 & 0.74$\pm$0.10 & 17.04$\pm$1.79 \\
         & \cellcolor{teal!15}\texttt{LLM-Tool Compiler} & \cellcolor{teal!15}0.46$\pm$0.09 & \cellcolor{teal!15}99.85$\pm$0.04 & \cellcolor{teal!15}94.73$\pm$0.18 & \cellcolor{teal!15}97.02$\pm$0.22 \\
        \hdashline
        \multirow{2}{*}{CoT - Few-Shot} & $^\dagger$Baseline: In-context + OAI Parallel &  2.84$\pm$0.11 & 99.00$\pm$0.02 & 0.25$\pm$0.01 & 36.01$\pm$0.45 \\
         & \cellcolor{teal!15}\texttt{LLM-Tool Compiler} & \cellcolor{teal!15}4.87$\pm$1.12 & \cellcolor{teal!15}99.90$\pm$0.02 & \cellcolor{teal!15}97.68$\pm$0.15 & \cellcolor{teal!15}99.41$\pm$0.04 \\
        \hdashline
        \multirow{2}{*}{ReAct - Zero-Shot} & $^\dagger$Baseline: In-context + OAI Parallel &  42.39$\pm$2.91 & 99.00$\pm$0.02 & 15.25$\pm$1.69 & 25.59$\pm$0.80 \\
         & \cellcolor{teal!15}\texttt{LLM-Tool Compiler} & \cellcolor{teal!15}40.00$\pm$3.06 & \cellcolor{teal!15}99.95$\pm$0.01 & \cellcolor{teal!15}79.68$\pm$2.00 & \cellcolor{teal!15}97.12$\pm$0.18 \\
        \hdashline
        \multirow{2}{*}{ReAct - Few-Shot} & $^\dagger$Baseline: In-context + OAI Parallel & 41.42$\pm$2.28 & 99.95$\pm$0.01 & 16.38$\pm$1.36 & 21.33$\pm$0.96 \\
         & \cellcolor{teal!15}\texttt{LLM-Tool Compiler} & \cellcolor{teal!15}51.93$\pm$4.52 & \cellcolor{teal!15}99.75$\pm$0.05 & \cellcolor{teal!15}93.30$\pm$1.74 & \cellcolor{teal!15}97.56$\pm$0.27 \\
        \midrule
        \multicolumn{2}{c}{Oracle} & \multicolumn{2}{c}{100.00\% (1971 Qs)} & \multicolumn{2}{c}{100.00 \% (4039 Qs)} \\
        \bottomrule
\end{tabular}
\end{table*}

\section{Results}

\begin{table}
  \caption{Breaking down the improvement for average time per task. We ablate the \texttt{LLM-Tool Compiler} results with and without concurrent tool execution, \textit{i.e.}, executing tools simultaneously as concurrent system processes.}
  \label{tab:concurrent_results}
  \begin{tabular}{lccc}
        \toprule
         & \multirow{2}{*}{Model} & Fused  & Fused+  \\
         &  & Only & Concurrent  \\
        \midrule
        \multirow{4}{*}{\textbf{\textit{GPT-3.5}}} & CoT Zero-Shot & 16.60 \% & 28.03 \% \\
        & CoT - Few-Shot & 14.35 \% & 26.27 \%  \\
        & ReAct - Zero-Shot & 22.50 \% & 37.05 \%  \\
        & ReAct - Few-Shot & 21.63 \% & 35.25 \% \\
        \midrule
        \multirow{4}{*}{\textbf{\textit{GPT-4}}} & CoT - Zero-Shot & 8.65 \% & 17.53 \%  \\
        & CoT - Few-Shot & 7.33 \% & 16.19 \%  \\
        & ReAct - Zero-Shot & 9.01 \% & 17.67 \%  \\
        & ReAct - Few-Shot & 8.74 \% & 17.45 \% \\
        \bottomrule
\end{tabular}
\end{table}

\textbf{Experimental Setup.} We assess the \texttt{LLM-Tool Compiler} on the \texttt{GeoLLM-Engine} benchmark~\cite{singh2024geoengine}, highlighting the effectiveness of our methodology on complex long-horizon many-tool remote sensing (RS) applications and satellite imagery. The benchmark, along with its associated engine, serves as a real-world testbed equipped with a comprehensive suite of open-source APIs, dynamic map/web UIs, as it supports various Python packages tailored for data analytics and geospatial tasks, including \texttt{RAG} and vector stores, \texttt{Mapbox} APIs for interactive mapping, and \texttt{GeoPandas} for geospatial data visualization and manipulation~\cite{singh2024geoqa}. Designed to handle LLMs across hundreds of API tools, the platform leverages a vast repository of geospatial data from prominent RS LLM benchmarks, encompassing over 5 million satellite images and hundreds of thousands of tasks meticulously organized into SQL tables for efficient querying and manipulation. This setup provides an ideal environment for implementing and evaluating the \texttt{LLM-Tool Compiler}.

\textbf{Metrics}. For LLM performance, we follow established agent evaluation practices~\cite{zhuang2023toolqa, maini2024tofu} and we compute the \textit{Success Rate} (the proportion of tasks successfully completed across the benchmark), the \textit{Correctness Ratio} (the proportion of correct function-call operations, noting that a wrong tool call might not prevent the agent from successfully completing the task), and the ROUGE-L score. We also report performance with respect to the underlying remote sensing tasks in the \texttt{GeoLLM-Engine} benchmark, with F1 and recall for object detection and land coverage classification (LCC), respectively, alongside ROUGE for visual question answering (VQA). In terms of system costs, we compute the average number of tokens per task and the average time per API tool call. To capture the total latency of serving API requests in cloud deployments, we maintain a running average across multiple GPTs, discarding outliers more than two standard deviations from the mean to ensure representative latency measurements. For compiler performance, we calculate the parallelization rate as the average number of tools being called "in parallel" per API call.

\subsection{Main results}
\label{subsec:main_results}

\textbf{Baselines}. We summarize agent peformance and our extensinve analysis in Tables~\ref{tab:results_agents} and \ref{tab:latency_results}. Across all our main experiments,  we run on the latest GPT releases (Turbo) with OpenAI's function calling API activated for both GPT-3.5 and GPT-4 \texttt{Turbo (0125)} versions. 
These releases are the top-performing OpenAI models in terms of function calling~\cite{berkeley-function-calling-leaderboard}, particularly for parallel execution. We consider two widely established prompting techniques, namely Chain-of-Thought (CoT)~\cite{wei2023chainofthought} and ReAct~\cite{yao2023react, yang2023mmreact}. While there are existing results previously reported on the \texttt{GeoLLM-Engine} benchmark for these methods~\cite{singh2024geoengine}, we aim to achieve the strongest baselines to the best of our abilities. 

To this end, we first investigate the original prompting from~\cite{singh2024geoengine} and observe partially incorrect function calls yet correct underlying logic (\textit{e.g.}, \texttt{filter\_loc} calls with arguments corresponding to the correct area but with latitude-longitude values not covering a wide enough range hence failing to provide good recall for satellite-based object detections). Recognizing room for improvement, we employed enhanced \textit{in-context} prompting to improve agent performance as much as possible, denoting this as ``$^\dagger$\textit{Baseline}.'' Table~\ref{tab:results_agents} shows that all our $^\dagger$\textit{Baseline} versions improve upon the original results reported in~\cite{singh2024geoengine}(\textit{Baseline}) across all models and metrics of interest, with success rates increasing by as much as 13\%. Therefore, it is important to note that instead of comparing to a rudimentary sequential ReAct baseline~\cite{kim2024compiler} assuming previously suboptimal metrics, we ensure that \texttt{LLM-Tool Compiler} is benchmarked against already highly competitive baselines.

\textbf{Improving agent effiency}. We present the average token per task and average time per task in Table \ref{tab:latency_results}. The \textit{Token Reduction} and \textit{Speedup} compared to the baseline are presented in separate columns in addition to the Success Rate. We observe that with negligible deduction in performance, \texttt{LLM-ToolCompiler} is able to reduce token cost by 1.3$\times$ to 1.4$\times$ in case of GPT-4 and 1.2$\times$ to 1.35$\times$ in case of GPT-3.5. This is consistently higher than OAI Parallel function calling which achieves token reduction ranging from 1.16$\times$ to 1.27$\times$ in case of GPT-4 and 1.01$\times$ to 1.15$\times$ in case of GPT-3.5. Interestingly, even with strong in-context prompting, OAI Parallel function calling is still able to provide some speedup for the tasks. The performance degradation for \texttt{LLM-ToolCompiler} is much less for GPT-4 (less that 1\%) as compared GPT-3.5 highlighting its ability for complex reasoning. We also observe that the speedup for \texttt{LLM-ToolCompiler} in terms of average time per task is higher than OAI Parallel function calling consistently. Interestingly even though token reduction in case of GPT-3.5 is relatively less than GPT-4, the speedup for the former is higher, this can be attributed majorly to low latency for GPT-3.5 API compared to GPT-4. There is no fixed pattern for improvement for different prompting techniques - CoT, ReAct or Few-Shot suggesting the improvement is majorly affected by the model being used.

\textbf{Agent performance}. We present the performance related metrics in Table \ref{tab:results_agents}. Despite the considerable improvements in latency discussed earlier, we observe only minor task-dependent degradation in agent performance. To provide a comprehensive overview, we include all metrics from the benchmark, and we find that our performance is well within the variance compared to OpenAI’s parallel-function calling. This degradation in performance is slightly more noticeable with GPT-3.5 than with GPT-4. Interestingly, when using GPT-4, most metrics show almost no degradation at all. In fact, we see improvement in some cases, especially for Correctness Rate. Compared to the baseline, the only two aspects exhibiting larger degradation are object detection (F1) and language metrics (Rouge-L), but these also display similar non-negligible drops for OpenAI’s parallel-function calling. This is primarily attributed to underlying task variance - for instance, runs that might miss certain prompts, such as failing to fully complete a remote-sensing task with granularity across an entire continent, would disproportionately affect overall recall due to the data point-intensive outcome of the completion. As this limitation has already been discussed in recent RS benchmark analyses \cite{singh2024geoqa}, we investigated this variance by repeating each method multiple times and confirmed wider variations for these metrics. Overall, any degradation observed remains within this variance range.

\subsection{Ablation studies}

\textbf{Benchmark parallelization analysis.}
As previously mentioned and inspired by prior research \cite{kim2024compiler}, conventional methodologies for parallel function calling often presume tasks with strong interdependencies, or they evaluate their approaches on custom parallelizable benchmarks (\textit{e.g.}, ParallelQA \cite{kim2024compiler}). In contrast, we build upon existing benchmarks that do not come with specific assumptions about parallelization, aiming to reflect real-world scenarios on large-scale Copilot systems. In this regard, we conduct an oracle analysis to explore the underlying benchmark's potential for overall parallelization achievable by an ideal agent. This analysis, while extending beyond the scope of our current methodology assessment, offers insights akin to an LLM equivalent of Amdahl’s law. Given that the majority of question types in \texttt{GeoLLM-Engine} involve operations like load operations (where loading different data sources can occur concurrently) and filtering operations (where different operations are suitable for multi-tool execution within a single API call), we focus this particular ablation on these. We define parallelization rate as the percentage of tasks that are parallelized compared to the total number of parallelizable tasks - that an oracle agent would do. 

In Table \ref{tab:parallelization_analysis}, we present our findings. Firstly, the choice of GPT version has a notable impact on performance. GPT-4 exhibits superior baseline performance compared to earlier versions, indicating optimization for Copilot tasks. For load ops, GPT-4 achieves near 100\% parallelization and for filter ops it achieves $>$ 97\%. Secondly, employing prompting techniques, particularly transitioning from chain-of-thought to ReAct, enhances parallelization rates for both load and filter operations. This improvement is attributed to the reasoning process facilitated by Thought-Reason-Act and the integration of few-shot examples, enabling concurrent execution of load operations. However, we also observe substantial room for improvement in baseline approaches. By leveraging a fuser, our method inherently achieves higher parallelization rates, surpassing baseline approaches by four to five times.

\begin{figure}[h]
  \centering
  \includegraphics[width=\linewidth]{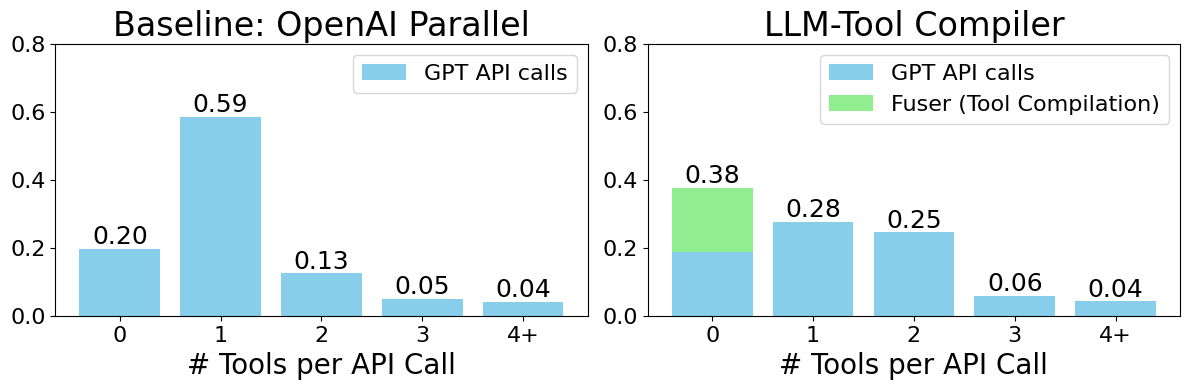}
  \caption{Distributions of the number of tool calls on the \texttt{GeoLLM-Engine-5k} for the baseline vs. \texttt{LLM-Tool Compiler} with GPT-4 ReAct - Zero-Shot prompting.}
  \Description{Number of Tools per API Call.}
  \label{fig:distro_tools}
\end{figure}

\textbf{Tool calls per step.} In this section we study the numnber of tools invoked per API call to GPT (step) to answer a user query. We compare the distributions of number of tools per step for c and OpenAI's parallel function calling baseline in Figure~\ref{fig:distro_tools}. We consider the GPT-4 ReAct - Zero-Shot prompting method both the parallel $^\dagger$Baseline and our  \texttt{LLM-Tool Compiler}. The 0-tools per call refers to the queries where no tools were required or when GPT returns the final answer after using required tools, as well as to the fuser executions since we do not execute any tools (Figure~\ref{fig:distro_tools}, right). We also see that the percentage of calls where 2 tools are used is significantly higher for \texttt{LLM-Tool Compiler} compared to OpenAI's parallel function calling (28\% vs 13\%) and percentage of calls with 1 tool per step is reduced for \texttt{LLM-Tool Compiler}. This shows the effectiveness of \texttt{LLM-Tool Compiler} for parallelizing and invoking more tools per API call.

\textbf{Efficiency modes}. In our methodology section, we clarified that the term `parallel' as used in OpenAI documentation, referring to multi-tool execution, does not imply simultaneous execution. The primary advantage, therefore, might initially stem from fewer API round-trips, with concurrent tool execution during each step further reducing latency. This analysis aims to determine the individual contributions of each component to the overall speed-up. We conducted tests with different methods, deactivating concurrent execution in our executor module to isolate the effects. In Table \ref{tab:concurrent_results}, we compare these modes and observe that the \textbf{fuser} module alone is able to able to bring the majority of the speedup implying efficient parallelization of tools bringing majority of the total speedup achieved in most cases. As expected, the concurrent execution in addition to the \textbf{fuser} module also contributes a significant jump in speed, in many cases doubling the speedup.

\textbf{Compiler GPT version}. Our analysis highlights a notable performance disparity between different GPT versions, as detailed in Table \ref{tab:results_agents}. It is well-established that GPT-4 outperforms earlier versions in function-calling tasks \cite{singh2024geoengine, singh2024geoqa}. However, the lower efficacy of our GPT-3.5 baselines may not solely stem from inherent limitations in function calls but also from sub-optimal tool selection by the fuser module. To explore this, we implemented a hybrid model for our analysis: while retaining GPT-4 for general API calls, we utilized GPT-3.5 exclusively for fuser operations. Results in Table \ref{tab:fuser_results} indicate that the performance of this hybrid model generally lies between the all-GPT-3.5 configuration and the all-GPT-4 setup, supporting our hypothesis that GPT-3.5's inferior function-calling capability is further exacerbated by compiler inefficiencies. As we extend our methodology to include open-source models like Llama 3\cite{meta-llama3}, this ablation study provides valuable insights into potential heterogeneous agent configurations, presenting exciting opportunities for hardware-related optimizations.

\begin{table}
  \caption{Ablating how \textit{fuser} performance (via different GPT version) affects the overall effectiveness of our approach.}
  \label{tab:fuser_results}
  \begin{tabular}{lcccccc}
        \toprule
        \multirow{2}{*}{Model} & \textit{fuser} & Success & Correct. & Token & Speedup \\
         & version & Rt (\%) $\uparrow$ & Rt (\%) $\uparrow$ & Rdc. $\uparrow$ & $\uparrow$ \\
        \midrule
        \multirow{2}{*}{CoT ZS} & \textbf{\textit{GPT-3.5}} & 72.31 & 86.40 & 1.47 $\times$ & 1.32 $\times$ \\
         & \cellcolor{teal!15}\textbf{\textit{GPT-4}} & \cellcolor{teal!15}84.15 & \cellcolor{teal!15}93.84 & \cellcolor{teal!15}1.39 $\times$ & \cellcolor{teal!15}1.21 $\times$ \\
        \hdashline
        \multirow{2}{*}{CoT FS} & \textbf{\textit{GPT-3.5}} & 70.15 & 86.04 & 1.51 $\times$ & 1.30 $\times$ \\
         & \cellcolor{teal!15}\textbf{\textit{GPT-4}} & \cellcolor{teal!15}82.51 & \cellcolor{teal!15}94.60 & \cellcolor{teal!15}1.41 $\times$ & \cellcolor{teal!15}1.19 $\times$ \\
        \hdashline
        \multirow{2}{*}{ReAct ZS} & \textbf{\textit{GPT-3.5}} & 69.49 & 85.10 & 1.44 $\times$ & 1.32 $\times$ \\
         & \cellcolor{teal!15}\textbf{\textit{GPT-4}} & \cellcolor{teal!15}81.68 & \cellcolor{teal!15}93.09 &\cellcolor{teal!15}1.34 $\times$ & \cellcolor{teal!15}1.21 $\times$ \\
        \hdashline
        \multirow{2}{*}{ReAct FS} & \textbf{\textit{GPT-3.5}} & 71.20 & 85.32 & 1.42 $\times$ & 1.32 $\times$ \\
         & \cellcolor{teal!15}\textbf{\textit{GPT-4}} & \cellcolor{teal!15}82.68 & \cellcolor{teal!15}94.11 & \cellcolor{teal!15}1.33 $\times$ & \cellcolor{teal!15}1.21 $\times$ \\
        \bottomrule
\end{tabular}
\end{table}

\textbf{Latency Analysis:} Motivated by the well-established practice in our community of hardware modeling that accurately captures hardware performance across various system metrics and components~\cite{marculescu2018hardware}, we are intrigued by the potential of such approaches in the context of LLMs, as also underscored in recent studies~\cite{yuan2024roof}. While it falls outside the immediate scope of our work, we would like to provide a preliminary analysis here, interested in studying the observation in~\cite{kim2024compiler}, that early stopping in function-calling could complicate accurate latency measurements.

\begin{figure}[ht!]
  \centering
  \includegraphics[width=\linewidth]{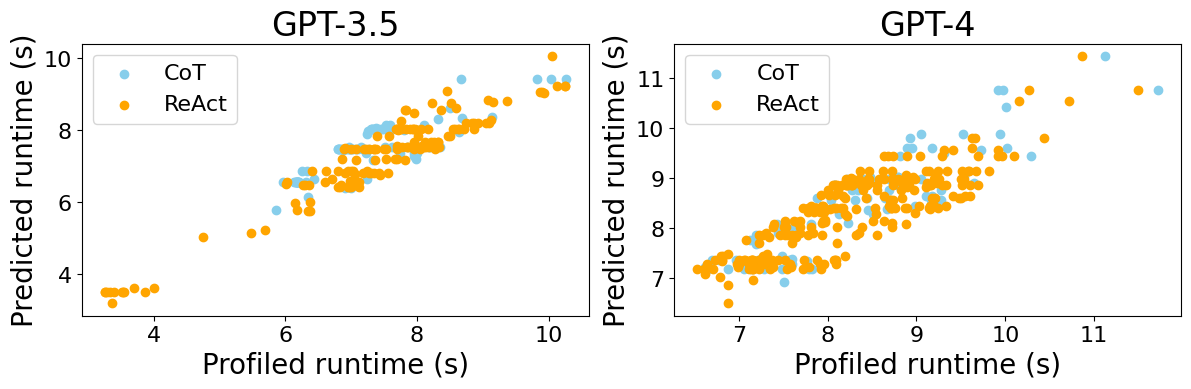}
  \caption{As a preliminary analysis, we capture whether a simple LUT-based modeling approach can capture overall agent latency. Shown below is the profiled and prediced average runtime per task across the different baselines considered in our experiments.}
  \Description{Assessing the LUT-based model.}
  \label{fig:runtime_model}
\end{figure}

We investigate whether a straightforward model assumption can effectively capture runtime in our experimental setup. We measure the average time per API tool call and per tool execution using the Python \texttt{time} package. To accurately assess total latency for API requests in cloud deployments, we maintain a running average across various endpoints and tool operations, excluding outliers that are more than two standard deviations from the mean. For these tests, we deploy hundreds of isolated GPT instances specifically for this evaluation to avoid congestion and allow for representative endpoint response times.

We construct a simple Look-up-Table (LUT) model where we record the average runtime for each tool and key API components (\textit{e.g.}, fuser call, final-answer API call, \textit{etc.}). Next, we randomly sample a fifth of the datapoints and rerun them, recording both the overall runtime and aggregating per-tool time based on the LUT entry. To normalize for the fact that different queries have different horizons, we divide the total prompt latency by the number of tools/calls in the agent solution path, and we report the predicted and actual runtime in Figure~\ref{fig:runtime_model} for both GPT versions and prompting techniques. While beyond the scope of our current work, we do observe that the LUT still provides a good approximation. Using the MSE-based metric from~\cite{cai2017neuralpower}, we observe that modeling error is 14.28\% and 17.46\% for the LUT-based heuristics. Of course, this is only a case-study, and the findings could be specific to the underlying platform. As part of our ongoing efforts and as we expand to local execution of open-source LLMs, we will be expanding this preliminary analysis to investigate where scalable and accurate runtime or power consumption modeling is feasible.

\section{Discussion and Future Work}

We acknowledge certain limitations in our present methodology. First, our system is implemented focusing on optimizing latency and bandwidth with extensive use of cloud endpoints. Building upon our initial findings, we aim to expand our analysis to local execution. Motivated by the relationship between runtime efficiency and hardware operational costs~\cite{balaskas2024hardware}, we anticipate that \texttt{LLM-Tool Compiler} could yield further energy and power consumption efficiencies. Towards enabling local execution, we will explore GPT alternatives that can be run locally, such as the recently intoduced Llama-3~\cite{meta-llama3} and Phi-3.5~\cite{abdin2024phi3} architectures. Given that approach is implemented in a nearly prompting-agnostic manner, we expect to flexibly incorporate it across different computational environments. On that note, we note other prompting schemes such as RAG~\cite{retrieval} and demonstration-retrieval~\cite{srinivasan2023nexusraven}, which we plan to explore in future work beyond CoT and ReAct.

Moreover, while we have not reported on other benchmarks, we plan to extend our evaluation beyond the geospatial domain to a wider range of orthogonal tasks also considered in recent system-level LLM optimization papers~\cite{kim2024compiler}. Last, we are currently investigating improving the \texttt{LLM-Tool Compiler} methodology further. For instance, one limiting aspect in the considered benchmark is that the order of operations does not matter. As an intriguing research direction, we aim to investigate graph-based approaches~\cite{karatzas2023omniboost} that capture dependencies and further optimize system performance.

\section{Conclusion}

In this work, we introduced \texttt{LLM-Tool Compiler}, a GPT-driven approach that dynamically identifies and fuses similar tool operations into unified tasks, inherently enhancing parallelization in function calling and reducing system latency. We demonstrated significant improvements over existing methods on a large-scale geospatial Copilot platform, achieving up to four times more parallel calls than existing methods while reducing token costs and latency by up to 40\% and 12\%, respectively. Our approach remains compatible with various established prompting techniques, offering a nearly agnostic solution that minimizes engineering overhead. By explicitly profiling and reporting hardware metrics alongside agent performance, \texttt{LLM-Tool Compiler} provides a comprehensive solution towards enhancing LLM efficiency on real-word multi-tool agent workloads.

\bibliographystyle{ACM-Reference-Format}
\bibliography{base}

\end{document}

%% file: tables.tex
\begin{table*}
  \caption{Agent performance metrics for \texttt{LLM-Tool Compiler} compared to baselines across various models and prompting techniques. First, we improve on the previously reported baselines in~\cite{singh2024geoengine}. Using the original prompting, we frequently observe incorrect function calls yet correct underlying logic (Section~\ref{subsec:main_results}). We incorporated in-context prompting to improve agent performance as much as possible, denoted as $^\dagger$Baseline. Baseline corresponds to the original results~\cite{singh2024geoengine} without our prompting.}
  \label{tab:results_agents}
  \begin{tabular}{lccccccc}
  \toprule
        \multirow{2}{*}{Model} & \multirow{2}{*}{Compiler} & Success & Correctness & Obj. Det & LCC & VQA  \\
        & & Rate (\%) $\uparrow$ & Rate (\%) $\uparrow$ & F1 (\%) $\uparrow$ & R $\% \uparrow$ & Rouge-L (\%) $\uparrow$ \\
        \midrule
        \textbf{\textit{GPT-3.5 Turbo (0125)}} & & & & & & & \\
\multirow{4}{*}{CoT - Zero-Shot} & Baseline~\cite{singh2024geoengine} & 64.66 & 40.20 & 54.99 & 93.28 & 54.40 \\
         & $^\dagger$Baseline: In-context & 65.73$\pm$0.45 & 69.22$\pm$0.34 & 59.94$\pm$7.55 & 96.73$\pm$4.85 & 54.55$\pm$0.07 \\
         & $^\dagger$Baseline: In-context + OAI Parallel & 65.93$\pm$0.49 & 69.68$\pm$0.18 & 55.22$\pm$24.98 & 93.28$\pm$26.61 & 54.50$\pm$0.16 \\
         & \cellcolor{teal!15}\texttt{LLM-Tool Compiler} & \cellcolor{teal!15}58.59$\pm$0.57 & \cellcolor{teal!15}66.88$\pm$0.54 & \cellcolor{teal!15}60.45$\pm$6.31 & \cellcolor{teal!15}79.04$\pm$3.34 & \cellcolor{teal!15}52.35$\pm$0.07 \\
        \hdashline
\multirow{4}{*}{CoT - Few-Shot}  & Baseline~\cite{singh2024geoengine} & 68.45 & 65.77 & 73.81 & 98.39 & 56.28  \\
         & $^\dagger$Baseline: In-context & 67.87$\pm$0.20 & 77.83$\pm$0.13 & 70.46$\pm$13.29 & 96.60$\pm$1.35 & 57.55$\pm$0.02 \\
         & $^\dagger$Baseline: In-context + OAI Parallel & 67.17$\pm$0.57 & 77.61$\pm$0.04 & 73.23$\pm$1.91 & 96.29$\pm$0.32 & 57.64$\pm$0.06 \\
         & \cellcolor{teal!15}\texttt{LLM-Tool Compiler} & \cellcolor{teal!15}59.55$\pm$0.20 & \cellcolor{teal!15}77.96$\pm$0.06 & \cellcolor{teal!15}59.81$\pm$16.52 & \cellcolor{teal!15}80.81$\pm$0.80 & \cellcolor{teal!15}54.65$\pm$0.07 \\
        \hdashline
\multirow{4}{*}{React - Zero-Shot}  & Baseline~\cite{singh2024geoengine} & 51.56 & 62.06 & 54.16 & 92.57 & 56.45 \\
         & $^\dagger$Baseline: In-context & 64.81$\pm$0.63 & 56.53$\pm$1.90 & 58.60$\pm$34.94 & 94.79$\pm$7.06 & 60.67$\pm$0.38 \\
         & $^\dagger$Baseline: In-context + OAI Parallel & 64.98$\pm$0.20 & 58.04$\pm$2.30 & 62.17$\pm$55.45 & 96.37$\pm$8.51 & 61.27$\pm$0.19 \\
         & \cellcolor{teal!15}\texttt{LLM-Tool Compiler} & \cellcolor{teal!15}59.33$\pm$0.67 & \cellcolor{teal!15}50.64$\pm$4.93 & \cellcolor{teal!15}51.86$\pm$62.10 & \cellcolor{teal!15}74.24$\pm$9.76 & \cellcolor{teal!15}57.46$\pm$0.14 \\
        \hdashline
\multirow{4}{*}{React - Few-Shot}  & Baseline~\cite{singh2024geoengine} & 73.47 & 68.42 & 75.01 & 97.45 & 65.26  \\
         & $^\dagger$Baseline: In-context & 73.59$\pm$0.13 & 80.15$\pm$0.34 & 67.84$\pm$11.52 & 97.34$\pm$4.13 & 65.57$\pm$0.02 \\
         & $^\dagger$Baseline: In-context + OAI Parallel & 73.20$\pm$0.19 & 79.65$\pm$0.16 & 70.04$\pm$19.57 & 97.51$\pm$0.49 & 65.32$\pm$0.01 \\
         & \cellcolor{teal!15}\texttt{LLM-Tool Compiler} & \cellcolor{teal!15}66.12$\pm$1.05 & \cellcolor{teal!15}73.63$\pm$1.65 & \cellcolor{teal!15}55.76$\pm$15.16 & \cellcolor{teal!15}86.82$\pm$5.97 & \cellcolor{teal!15}61.49$\pm$0.35 \\
        \midrule
        \textbf{\textit{GPT-4 Turbo (0125)}} & & & & & & & \\
\multirow{4}{*}{CoT - Zero-Shot}  & Baseline~\cite{singh2024geoengine} & 77.35 & 80.88 & 87.99 & 96.56 & 65.29  \\
         & $^\dagger$Baseline: In-context & 84.39$\pm$0.14 & 93.25$\pm$0.09 & 80.60$\pm$0.14 & 95.98$\pm$0.73 & 63.79$\pm$0.08 \\
         & $^\dagger$Baseline: In-context+ OAI Parallel & 84.73$\pm$0.21 & 93.37$\pm$0.05 & 80.98$\pm$0.07 & 95.67$\pm$1.43 & 63.86$\pm$0.11 \\
         & \cellcolor{teal!15}\texttt{LLM-Tool Compiler} & \cellcolor{teal!15}84.15$\pm$0.20 & \cellcolor{teal!15}93.84$\pm$0.17 & \cellcolor{teal!15}74.16$\pm$0.08 & \cellcolor{teal!15}96.68$\pm$3.20 & \cellcolor{teal!15}61.37$\pm$0.08 \\
        \hdashline
\multirow{4}{*}{CoT - Few-Shot}  & Baseline~\cite{singh2024geoengine} & 80.00 & 84.01 & 88.40 & 99.89 & 67.65 \\
         & $^\dagger$Baseline: In-context & 83.15$\pm$0.22 & 93.45$\pm$0.14 & 75.04$\pm$22.23 & 97.53$\pm$0.51 & 61.80$\pm$0.06 \\
         & $^\dagger$Baseline: In-context + OAI Parallel & 83.05$\pm$0.46 & 93.30$\pm$0.05 & 74.21$\pm$23.21 & 98.17$\pm$0.32 & 62.16$\pm$0.17 \\
         & \cellcolor{teal!15}\texttt{LLM-Tool Compiler} & \cellcolor{teal!15}82.51$\pm$0.03 & \cellcolor{teal!15}94.60$\pm$0.07 & \cellcolor{teal!15}75.48$\pm$9.71 & \cellcolor{teal!15}95.51$\pm$0.88 & \cellcolor{teal!15}59.87$\pm$0.17 \\
        \hdashline
\multirow{4}{*}{React - Zero-Shot} & Baseline~\cite{singh2024geoengine} & 80.03 & 84.27 & 89.34 & 98.83 & 68.11 \\
         & $^\dagger$Baseline: In-context & 82.02$\pm$0.27 & 92.61$\pm$0.06 & 79.16$\pm$24.45 & 96.23$\pm$0.75 & 65.03$\pm$0.07 \\
         & $^\dagger$Baseline: In-context + OAI Parallel & 82.31$\pm$1.04 & 92.69$\pm$0.03 & 81.26$\pm$0.05 & 95.66$\pm$0.74 & 65.07$\pm$0.02 \\
         & \cellcolor{teal!15}\texttt{LLM-Tool Compiler} & \cellcolor{teal!15}81.68$\pm$0.31 & \cellcolor{teal!15}93.09$\pm$0.07 & \cellcolor{teal!15}72.92$\pm$1.76 & \cellcolor{teal!15}95.41$\pm$1.56 & \cellcolor{teal!15}64.44$\pm$0.03 \\
        \hdashline
\multirow{4}{*}{React - Few-Shot} & Baseline~\cite{singh2024geoengine} & 81.11 & 84.31 & 83.85 & 99.63 & 69.37   \\
         & $^\dagger$Baseline: In-context & 84.26$\pm$0.29 & 92.74$\pm$0.03 & 75.98$\pm$12.74 & 95.49$\pm$8.17 & 66.54$\pm$0.04 \\
         & $^\dagger$Baseline: In-context + OAI Parallel & 84.29$\pm$0.42 & 92.45$\pm$0.02 & 80.20$\pm$6.52 & 96.91$\pm$1.79 & 66.34$\pm$0.04 \\
         & \cellcolor{teal!15}\texttt{LLM-Tool Compiler} & \cellcolor{teal!15}82.68$\pm$0.27 & \cellcolor{teal!15}94.11$\pm$0.02 & \cellcolor{teal!15}76.77$\pm$10.34 & \cellcolor{teal!15}95.59$\pm$1.72 & \cellcolor{teal!15}65.48$\pm$0.02 \\
        \bottomrule
\end{tabular}
\end{table*}